# On Utilization and Importance of Perl Status Reporter (SRr) in Text Mining

Sugam Sharma[1], Tzusheng Pei[2], and Hari Cohly[2]

Department of Computer Science, Iowa State University[1], USA
Center for Bioinformatics, Jackson State University[2], USA
*sugam.k.sharma@gmail.com*

**ABSTRACT**

*In Bioinformatics, text mining and text data mining sometimes interchangeably used is a process to derive high-quality information from text. Perl Status Reporter (SRr) [1] is a data fetching tool from a flat text file and in this research paper we illustrate the use of SRr in text/data mining. SRr needs a flat text input file where the mining process to be performed. SRr reads input file and derives the high-quality information from it. Typically text mining tasks are text categorization, text clustering, concept and entity extraction, and document summarization. SRr can be utilized for any of these tasks with little or none customizing efforts. In our implementation we perform text categorization mining operation on input file. The input file has two parameters of interest (firstKey and secondKey). The composition of these two parameters describes the uniqueness of entries in that file in the similar manner as done by composite key in database. SRr reads the input file line by line and extracts the parameters of interest and form a composite key by joining them together. It subsequently generates an output file consisting of the name as firstKey_secondKey. SRr reads the input file and tracks the composite key. It further stores all that data lines, having the same composite key, in output file generated by SRr based on that composite key.*

## 1.0 INTRODUCTION

Labour-intensive manual text-mining approaches first came into picture in the mid-1980s, but for past decades technological advances have dominated the field with fast pace. As most of the information is stored in text, Text mining is blowing with a high demand.

Text mining is fundamental in extracting knowledge from growing literature in different research areas such as Bioinformatics. Text mining techniques are used to extract useful information and knowledge from unstructured text and different databases such as pubmed [2]. We apply Perl SRr on traffic database for texting mining to analyze the traffic data. The input file is a flat file in textual format contains traffic data. The first column entry is the identification (Id) of the city where the traffic has been collected from. This Id is unique for a particular metro city. The second column entry is the name of that metro city. The third column entry is the name of the company responsible for data collection in the respective metro city area. Next column entry is the date, the traffic data has been collected on. Rest of the column entries are self informative and are not much important in our development. Column entries first and third are of our interest and we call them as parameters of interest.

In this paper we call the first column enrty as firstKey and third column entry as secondKey. As text /data mining is somewhere concern with database so we try to relate our parameters of interest with composite key concept of database management system. In text mining there is a key factor which is the base for information retrieval. Likewise in our implementation firstKey and secondKey are the main factors in mining. SRr joins these keys together and form a composite key, firstKey_secondKey. SRr uses this composite key to segregate the data from the input data file.

SRr reads the input text file one line at a time, eradicates its separator, and stores the contents of that line in an array. The filled array is further operated to extract firstKey and secondKey. SRr joins these keys together and form a composite key subsequently. SRr keeps tracking the input data and generates output files based on the composite key. The name of the file will be same as the composite key. The extracted data consisting of the same composite key will be stored in that generated output file. Thus for an individual composite key, SRs generates a separate output file and fill it in the same fashion as stated above.

The rest of the paper is organized as follows.

Section 2.0 is the mining process section. This section details about the recent mining trends in Computer Science and Bioinformatics, and mining operation SRr performs on input text file. Section 3.0 is the result section. This section consists of input sample file we use in our implementation and sample output produced. Section 4.0 is the conclusion section and the last section 5.0 is the reference section.

## 2.0 MINING PROCESS AND SRr MINING

Text mining is utmost popular under Bioinformatics umbrella and has a big range of text mining applications in the biomedical literature. Mining can be described as discovery by computer of new, previously unknown information, by automatically extracting information from different written resources. In our development we use a flat file in textual format. A key element is to utilize the extracted information to form new facts or new hypotheses to be explored further by more conventional means of experimentation.

Text mining is different from web search. In search, the user has a goal to search for something that is already written by some one and exists in the database. In text mining, the target is to discover unknown information that is not yet known and could not have yet written down. Text mining is a variation on a field called data mining that tries to find interesting patterns from large databases.

There are many upcoming tools being developed every day for text mining and Perl SRr is one more development in the same direction and for the same purpose. The beauty of SRr over existing tool is that SRr is completely written in Perl and is highly customizable. The scope of operation of SRr is not only limited to text mining in Bioinformatics but also text mining in other areas as well where SRr is capable to perform mining process with more than 95% accuracy. SRr is a multipurpose tool, well efficient in information extraction process.

SRr performs the mining process on flat text input file. The following code snippet describes how SRr reads the input file and parses the sentences and categories the text. We write the sample code block-wise to make easy understanding.

In the following segment we declare the variables globally. To *$prc_file* variable we assign the input file to be processed. In our implementation we use text_mining.txt as the input file and is a flat textual file.

```
my $prc_file="text_mining.txt";
my $output1_file;
my $first_key;
my $store_first_key;
my $city_Name;
my $second_key;
my $store_second_key;
```

In following block of code SRr reads a file using a file handler INPUTHANDLER. If SRr is unable to open the input file, it throws a message.

```
open INPUTHANDLER, "<$prc_file" || die
"Could not open file for reading.\n";
print"\n SRr has performed the text mining
operations on input file \n";
```

In the following segment SRr reads the input flat file one line at a time, splits the line based on the separator used and stores the contents in an array.

```
while (<INPUTHANDLER>) {
my@cur_line = split /,/, $_ ;
my @store_cur_line = @cur_line;
```

Now SRr extracts the text from current line. We assign the parameters of interest in *$first_key* and *$second_key* variables respectively. The following block of code depicts this.

```
$first_key = shift @cur_line;
$city_Name = shift @cur_line;
$second_key = shift @cur_line;
```

The following segment is bit confusing but not a rocket science definitely. This part of SRr just tracks the composite attributes, generates the file based on those attributes and stores those data line containing the same composite attributes in that file. SRr uses if-els statement to separate out the generation of different files. The generated files are opened in append mode and any number of line can be added to it. SRr uses a file handler, OUTPUTHANDLER to write out the data to the file.

```
if($store_first_key eq $first_key &&
$store_second_key eq $second_key){

output1_file = "$first_key"."_"."$second_key".".txt";
open OUTPUTHANDLER, ">>$output1_file" ||
        die "Could not open file for writing.\n";
print OUTPUTHANDLER " @store_cur_line\n";

}
 elsif($store_first_key eq $first_key &&
$store_second_key ne $second_key){

$output1_file="$first_key"."_"."$second_key".".txt";
open OUTPUTHANDLER, ">>$output1_file" ||
            die"Couldnot  open file for writing.\n";
print OUTPUTHANDLER " @store_cur_line\n";

}

 elsif($store_first_key ne $first_key &&
$store_second_key eq $second_key){

$output1_file="$first_key"."_"."$second_key".".txt";
open OUTPUTHANDLER, ">>$output1_file" ||
        die "Could not open file for writing.\n";
print OUTPUTHANDLER " @store_cur_line\n";

}

 else {
$output1_file="$first_key"."_"."$second_key".".txt";
open OUTPUTHANDLER, ">>$output1_file" ||
```

```
    die "Could not open file for writing.\n";
print OUTPUTHANDLER " @store_cur_line\n";

}
$store_first_key = $first_key;
$store_second_key  = $second_key;
}
```

## 3.0 RESULTS

### 3.1 Input file snippet

In this section we show the curtailed sample input text file we use in our implementation. The data in the file is road traffic data in different metro cities in United States. The first column represents an Id of the city, and second column represents the name of the city, where the traffic belongs to. The third column represents the company name which collects the traffic data for that city. Rests of the columns are informative enough to understand. Column first and third are of our interest and together they form a composite key which differentiates the line from other lines.

```
7,NewYork,TRANSCOM,11/27/2008,23:55,TRANSCOM,1
1254,59571,No road selected,0,,W,THRU,5,5,,5,54,5,,5,,,,
7,NewYork,TRANSCOM,11/27/2008,23:55,TRANSCOM,1
1255,59572,No road selected,0,,N,THRU,5,5,,5,41,5,,5,,,,
7,NewYork,TRANSCOM,11/27/2008,23:55,TRANSCOM,1
1256,59573,No road selected,0,,S,THRU,5,5,,5,6,5,,5,,,,
7,NewYork,TRANSCOM,11/27/2008,23:55,TRANSCOM,2
5233,74179,No road selected,0,,E,THRU,5,5,,5,49,5,,5,,,,
7,NewYork,TRANSCOM,11/27/2008,23:55,TRANSCOM,2
5252,74191,No road selected,0,,E,THRU,5,5,,5,45,5,,5,,,,
16,Chicago,INDOT,11/27/2008,00:00,INDOT,31514,77443,
No road selected,1,LEFT,W,THRU,5,4,14,4,77,4,1.5,4,14,,,
16,Chicago,INDOT,11/27/2008,00:00,INDOT,31514,77443,
No road selected,2,LEFT
CENTER,W,THRU,5,4,25,4,72,4,4.3,4,25,,,
16,Chicago,INDOT,11/27/2008,00:00,INDOT,31514,77443,
No road selected,3,RIGHT
CENTER,W,THRU,5,4,36,4,56,4,11,4,36,,,
16,Chicago,INDOT,11/27/2008,00:00,INDOT,31514,77443,
No road selected,4,RIGHT,W,THRU,5,4,7,4,74,4,1,4,7,,,
16,Chicago,INDOT,11/27/2008,00:00,INDOT,31515,77444,
No road selected,1,LEFT,E,THRU,5,4,5,4,42,4,1,4,5,,,
16,Chicago,INDOT,11/27/2008,00:00,INDOT,31515,77444,
No road selected,2,LEFT
CENTER,E,THRU,5,4,21,4,66,4,4,4,21,,,
16,Chicago,INDOT,11/27/2008,00:00,INDOT,31515,77444,
No road selected,3,RIGHT
CENTER,E,THRU,5,4,12,4,72,4,3.8,4,12,,,
16,Chicago,INDOT,11/27/2008,00:00,INDOT,31515,77444,
No road selected,4,RIGHT,E,THRU,5,4,17,4,67,4,1.8,4,17,,,
16,Chicago,INDOT,11/27/2008,00:00,INDOT,31516,77445,
No road selected,1,LEFT,E,THRU,5,4,7,4,75,4,.5,4,7,,,
16,Chicago,INDOT,11/27/2008,00:00,INDOT,31516,77445,
No road
selected,2,CENTER,E,THRU,5,4,20,4,52,4,7.5,4,20,,,
16,Chicago,INDOT,11/27/2008,00:00,INDOT,31516,77445,
No road selected,3,RIGHT,W,THRU,5,4,20,4,54,4,4.8,4,20,,,
16,Chicago,MT,11/27/2008,00:10,MT,25040,73993,I-
355,1,LEFT,N,THRU,5,5,20,5,53,5,2.6,5,14,4,2,0
16,Chicago,MT,11/27/2008,00:10,MT,25040,73993,I-
355,2,CENTER,N,THRU,5,5,13,5,49,5,1.2,5,12,0,1,0
16,Chicago,MT,11/27/2008,00:10,MT,25040,73993,I-
355,3,RIGHT,N,THRU,5,5,5,5,60,3,.3,5,5,0,0,0
16,Chicago,MT,11/27/2008,00:10,MT,25040,73993,I-
355,4,LEFT,S,THRU,5,5,7,5,57,4,.7,5,5,2,0,0
16,Chicago,MT,11/27/2008,00:10,MT,25040,73993,I-
355,5,CENTER,S,THRU,5,5,14,5,63,4,1.2,5,13,0,1,0
16,Chicago,MT,11/27/2008,00:10,MT,25040,73993,I-
355,6,RIGHT,S,THRU,5,5,7,5,56,4,.4,5,7,0,0,0
16,Chicago,MT,11/27/2008,00:10,MT,25041,73994,I-
355,1,LEFT,N,THRU,5,5,15,5,61,5,2,5,9,4,2,0
16,Chicago,MT,11/27/2008,00:10,MT,25041,73994,I-
355,2,LEFT CENTER,N,THRU,5,5,9,5,68,4,.7,5,8,0,1,0
17,Baltimore,MDSHA,11/27/2008,00:00,MDSHA,18646,67
685,I-795,1,,S,THRU,1,1,12,1,63,1,,0,12,,,
17,Baltimore,MDSHA,11/27/2008,00:00,MDSHA,18648,67
687,I-83,1,,S,THRU,1,1,28,1,61,1,,0,28,,,
17,Baltimore,MDSHA,11/27/2008,00:00,MDSHA,18649,67
688,RT-32,1,,E,THRU,1,1,20,1,63,1,1,1,20,,,
17,Baltimore,MDSHA,11/27/2008,00:00,MDSHA,18651,67
690,I-795,1,,S,THRU,1,1,42,1,70,1,1,1,42,,,
17,Baltimore,MDSHA,11/27/2008,00:00,MDSHA,18652,67
691,I-70,1,,E,THRU,1,1,19,1,60,1,,0,19,,,
17,Baltimore,MDSHA,11/27/2008,00:00,MDSHA,18654,67
693,I-97,1,,S,THRU,1,1,10,1,70,1,,0,10,,,
17,Baltimore,MDSHA,11/27/2008,00:00,MDSHA,18655,67
694,I-695,1,,S,THRU,1,1,108,1,66,1,1,1,108,,,
17,Baltimore,MDSHA,11/27/2008,00:00,MDSHA,18658,67
697,I-795,1,,S,THRU,1,1,22,1,64,1,1,1,22,,,
17,Baltimore,MDSHA,11/27/2008,00:00,MDSHA,18659,67
698,I-83,1,,N,THRU,1,1,50,1,67,1,1,1,50,,,
17,Baltimore,MT,11/27/2008,00:00,MDSHA,18646,67685,I-
795,1,,S,THRU,1,1,12,1,63,1,,0,12,,,
17,Baltimore,MT,11/27/2008,00:00,MDSHA,18648,67687,I-
83,1,,S,THRU,1,1,28,1,61,1,,0,28,,,
17,Baltimore,MT,11/27/2008,00:00,MDSHA,18649,67688,R
T-32,1,,E,THRU,1,1,20,1,63,1,1,1,20,,,
17,Baltimore,MT,11/27/2008,00:00,MDSHA,18651,67690,I-
795,1,,S,THRU,1,1,42,1,70,1,1,1,42,,,
17,Baltimore,MT,11/27/2008,00:00,MDSHA,18652,67691,I-
70,1,,E,THRU,1,1,19,1,60,1,,0,19,,,
17,Baltimore,MT,11/27/2008,00:00,MDSHA,18654,67693,I-
97,1,,S,THRU,1,1,10,1,70,1,,0,10,,,
17,Baltimore,MT,11/27/2008,00:00,MDSHA,18655,67694,I-
695,1,,S,THRU,1,1,108,1,66,1,1,1,108,,,
17,Baltimore,MT,11/27/2008,00:00,MDSHA,18658,67697,I-
795,1,,S,THRU,1,1,22,1,64,1,1,1,22,,,
1,Pittsburgh,MT,11/27/2008,00:00,MT,7010,38,PA-
910,1,RIGHT,E,THRU,5,5,7,5,57,4,.8,5,7,,0,
1,Pittsburgh,MT,11/27/2008,00:00,MT,7010,38,PA-
910,2,LEFT,E,THRU,5,5,3,5,41,3,.6,5,3,,0,
1,Pittsburgh,MT,11/27/2008,00:00,MT,7010,38,PA-
910,3,RIGHT,W,THRU,5,5,3,5,56,2,.4,5,3,,0,
1,Pittsburgh,MT,11/27/2008,00:00,MT,7010,38,PA-
910,4,LEFT,W,THRU,5,5,5,5,46,2,.4,5,5,,0,
```

1,Pittsburgh,MT,11/27/2008,00:00,MT,7012,42,I-79,1,RIGHT,N,THRU,5,5,14,5,69,2,.8,5,12,,2,
1,Pittsburgh,MT,11/27/2008,00:00,MT,7012,42,I-79,2,CENTER,N,THRU,5,5,44,5,58,5,2.2,5,42,,2,
1,Pittsburgh,MT,11/27/2008,00:00,MT,7012,42,I-79,3,LEFT,N,THRU,5,5,10,5,58,5,1,5,10,,0,
1,Pittsburgh,MT,11/27/2008,00:00,MT,7012,42,I-79,4,RIGHT,S,THRU,5,5,8,5,19,2,1,5,8,,0,
1,Pittsburgh,MT,11/27/2008,00:00,MT,7012,42,I-79,5,CENTER,S,THRU,5,5,20,5,56,1,1.2,5,17,,3,
1,Pittsburgh,IDOT,11/27/2008,00:00,MT,7010,38,PA-910,1,RIGHT,E,THRU,5,5,7,5,57,4,.8,5,7,,0,
1,Pittsburgh,IDOT,11/27/2008,00:00,MT,7010,38,PA-910,2,LEFT,E,THRU,5,5,3,5,41,3,.6,5,3,,0,
1,Pittsburgh,IDOT,11/27/2008,00:00,MT,7010,38,PA-910,3,RIGHT,W,THRU,5,5,3,5,56,2,.4,5,3,,0,

### 3.2 Output file Snapshot

In this section we collect the results. We capture snapshots of all output files.

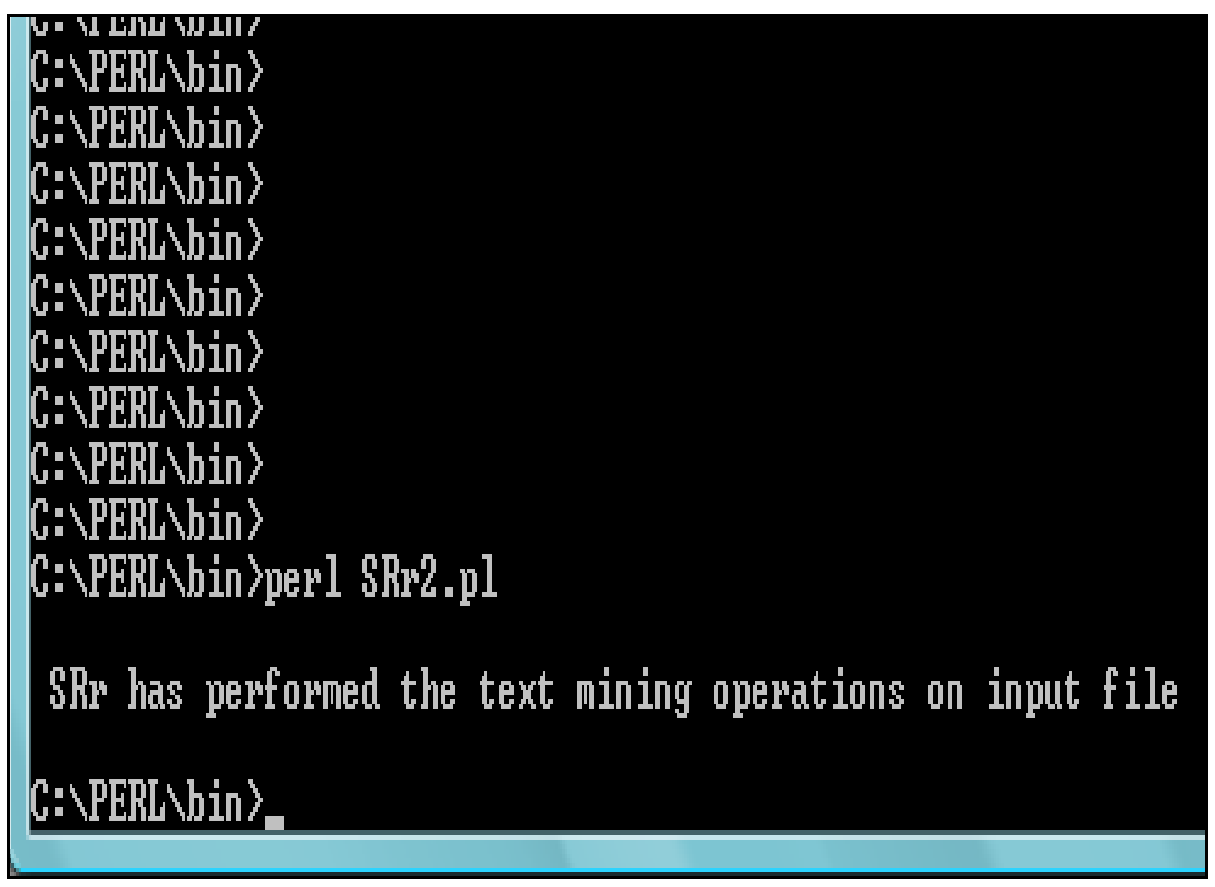
C:\PERL\bin>
C:\PERL\bin>
C:\PERL\bin>
C:\PERL\bin>
C:\PERL\bin>
C:\PERL\bin>
C:\PERL\bin>
C:\PERL\bin>
C:\PERL\bin>
C:\PERL\bin>perl SRr2.pl

SRr has performed the text mining operations on input file

C:\PERL\bin>_

Figure 1. SRr Status Screen Display

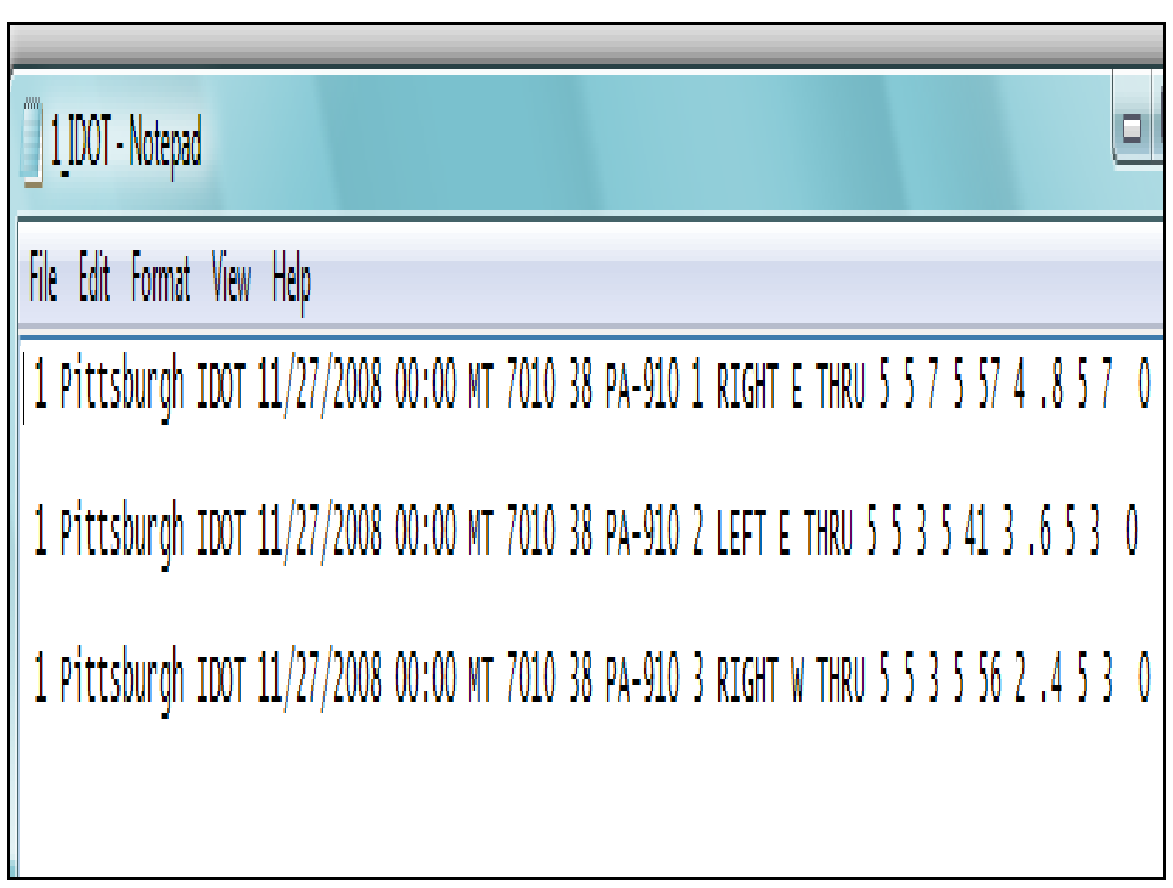
1_IDOT - Notepad
File Edit Format View Help
1 Pittsburgh IDOT 11/27/2008 00:00 MT 7010 38 PA-910 1 RIGHT E THRU 5 5 7 5 57 4 .8 5 7 0
1 Pittsburgh IDOT 11/27/2008 00:00 MT 7010 38 PA-910 2 LEFT E THRU 5 5 3 5 41 3 .6 5 3 0
1 Pittsburgh IDOT 11/27/2008 00:00 MT 7010 38 PA-910 3 RIGHT W THRU 5 5 3 5 56 2 .4 5 3 0

Figure 2. Extracted text stored in generated file 1_IDOT

input file which has 1 in first column and IDOT in third column and generates a file with name 1_IDOT before storing the extracted text. Figure 3 shows the stored data in the file 1_MT, generated by SRr. Figure 4 shows the contents of the file 7_TRANSCOM. SRr generates a file named as 16_INDOT. Figure 5 shows the contents of that file. Figure 6 shows the generated file 16_MT and its contents. Figures 7 represents file 17_MDSHA and the data stored in that. Figure 8 is 17_MT file showing its contents extracted from the input file and stored by SRr.

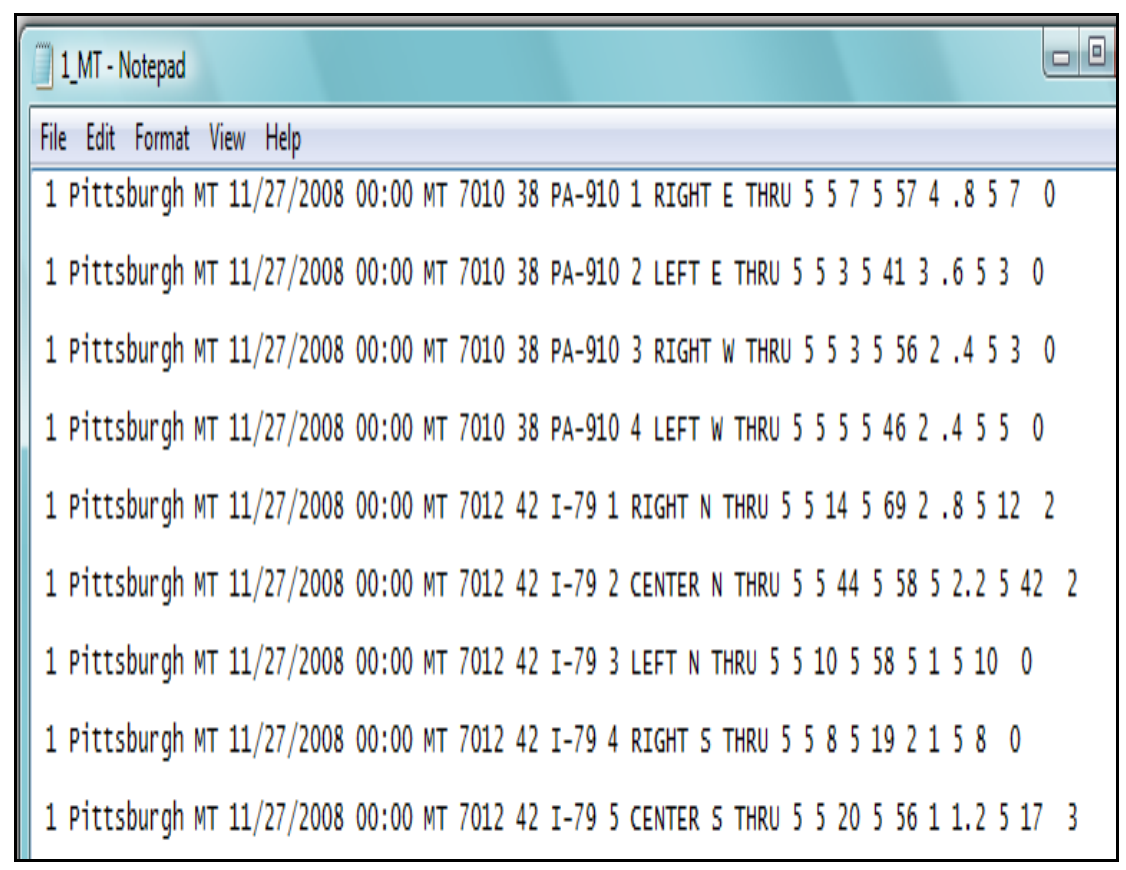
1_MT - Notepad
File Edit Format View Help
1 Pittsburgh MT 11/27/2008 00:00 MT 7010 38 PA-910 1 RIGHT E THRU 5 5 7 5 57 4 .8 5 7 0
1 Pittsburgh MT 11/27/2008 00:00 MT 7010 38 PA-910 2 LEFT E THRU 5 5 3 5 41 3 .6 5 3 0
1 Pittsburgh MT 11/27/2008 00:00 MT 7010 38 PA-910 3 RIGHT W THRU 5 5 3 5 56 2 .4 5 3 0
1 Pittsburgh MT 11/27/2008 00:00 MT 7010 38 PA-910 4 LEFT W THRU 5 5 5 5 46 2 .4 5 5 0
1 Pittsburgh MT 11/27/2008 00:00 MT 7012 42 I-79 1 RIGHT N THRU 5 5 14 5 69 2 .8 5 12 2
1 Pittsburgh MT 11/27/2008 00:00 MT 7012 42 I-79 2 CENTER N THRU 5 5 44 5 58 5 2.2 5 42 2
1 Pittsburgh MT 11/27/2008 00:00 MT 7012 42 I-79 3 LEFT N THRU 5 5 10 5 58 5 1 5 10 0
1 Pittsburgh MT 11/27/2008 00:00 MT 7012 42 I-79 4 RIGHT S THRU 5 5 8 5 19 2 1 5 8 0
1 Pittsburgh MT 11/27/2008 00:00 MT 7012 42 I-79 5 CENTER S THRU 5 5 20 5 56 1 1.2 5 17 3

Figure 3. Extracted text stored in generated file 1_MT

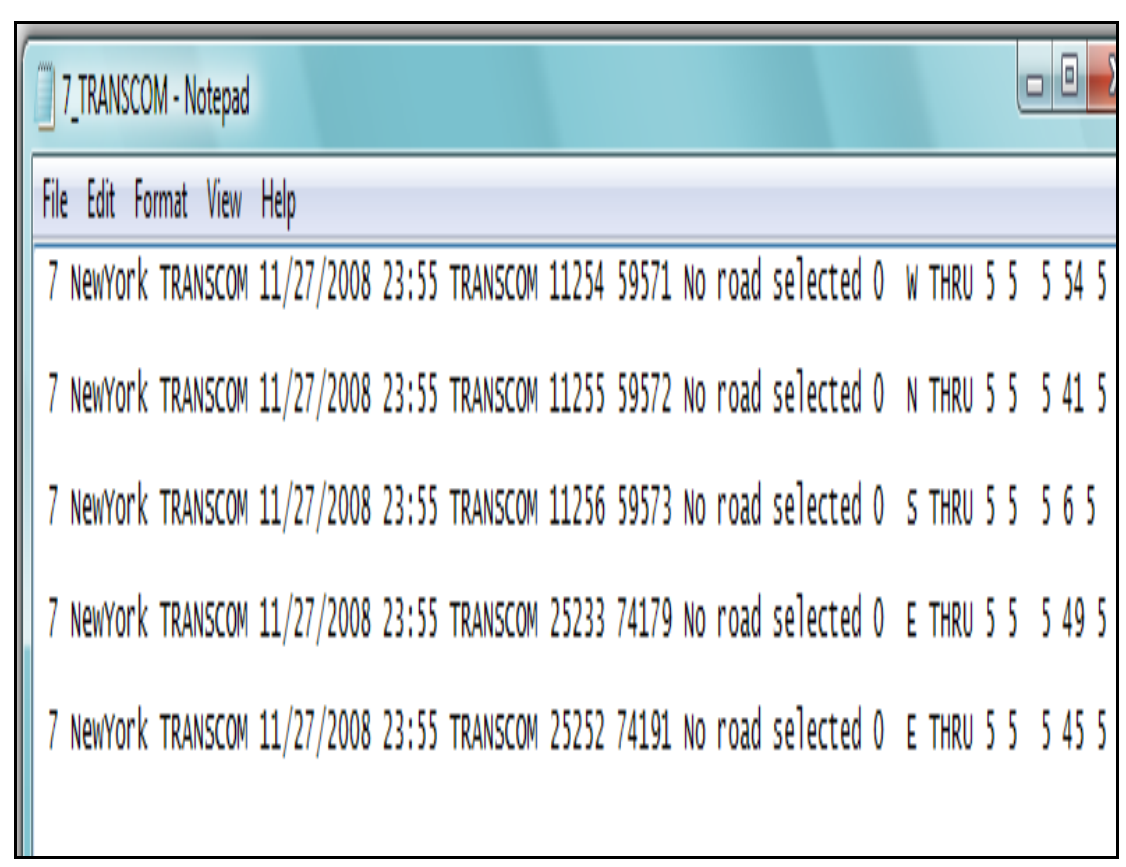
7_TRANSCOM - Notepad
File Edit Format View Help
7 NewYork TRANSCOM 11/27/2008 23:55 TRANSCOM 11254 59571 No road selected 0 W THRU 5 5 5 54 5
7 NewYork TRANSCOM 11/27/2008 23:55 TRANSCOM 11255 59572 No road selected 0 N THRU 5 5 5 41 5
7 NewYork TRANSCOM 11/27/2008 23:55 TRANSCOM 11256 59573 No road selected 0 S THRU 5 5 5 6 5
7 NewYork TRANSCOM 11/27/2008 23:55 TRANSCOM 25233 74179 No road selected 0 E THRU 5 5 5 49 5
7 NewYork TRANSCOM 11/27/2008 23:55 TRANSCOM 25252 74191 No road selected 0 E THRU 5 5 5 45 5

Figure 4. Extracted text stored in generated file 7_TRANSCOM

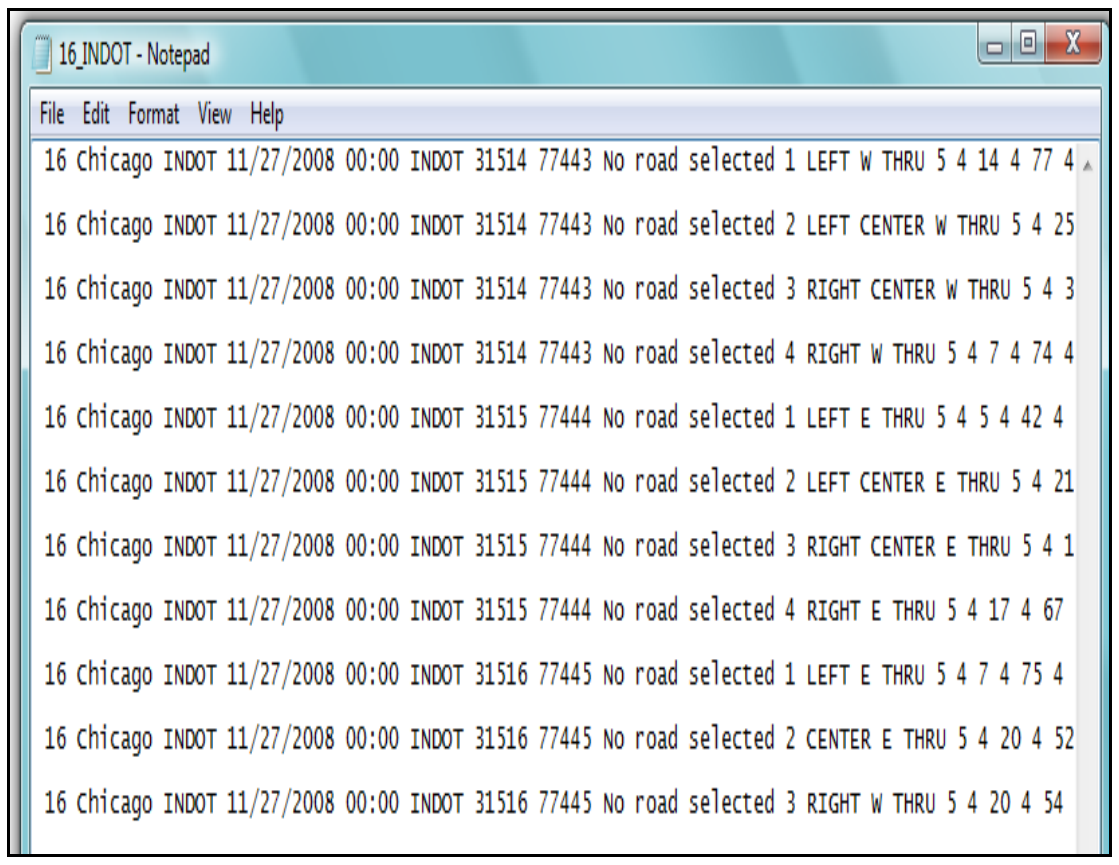

```
16_INDOT - Notepad
File Edit Format View Help
16 Chicago INDOT 11/27/2008 00:00 INDOT 31514 77443 No road selected 1 LEFT W THRU 5 4 14 4 77 4
16 Chicago INDOT 11/27/2008 00:00 INDOT 31514 77443 No road selected 2 LEFT CENTER W THRU 5 4 25
16 Chicago INDOT 11/27/2008 00:00 INDOT 31514 77443 No road selected 3 RIGHT CENTER W THRU 5 4 3
16 Chicago INDOT 11/27/2008 00:00 INDOT 31514 77443 No road selected 4 RIGHT W THRU 5 4 7 4 74 4
16 Chicago INDOT 11/27/2008 00:00 INDOT 31515 77444 No road selected 1 LEFT E THRU 5 4 5 4 42 4
16 Chicago INDOT 11/27/2008 00:00 INDOT 31515 77444 No road selected 2 LEFT CENTER E THRU 5 4 21
16 Chicago INDOT 11/27/2008 00:00 INDOT 31515 77444 No road selected 3 RIGHT CENTER E THRU 5 4 1
16 Chicago INDOT 11/27/2008 00:00 INDOT 31515 77444 No road selected 4 RIGHT E THRU 5 4 17 4 67
16 Chicago INDOT 11/27/2008 00:00 INDOT 31516 77445 No road selected 1 LEFT E THRU 5 4 7 4 75 4
16 Chicago INDOT 11/27/2008 00:00 INDOT 31516 77445 No road selected 2 CENTER E THRU 5 4 20 4 52
16 Chicago INDOT 11/27/2008 00:00 INDOT 31516 77445 No road selected 3 RIGHT W THRU 5 4 20 4 54
```

Figure 5. Extracted text stored in generated file 16_INDOT

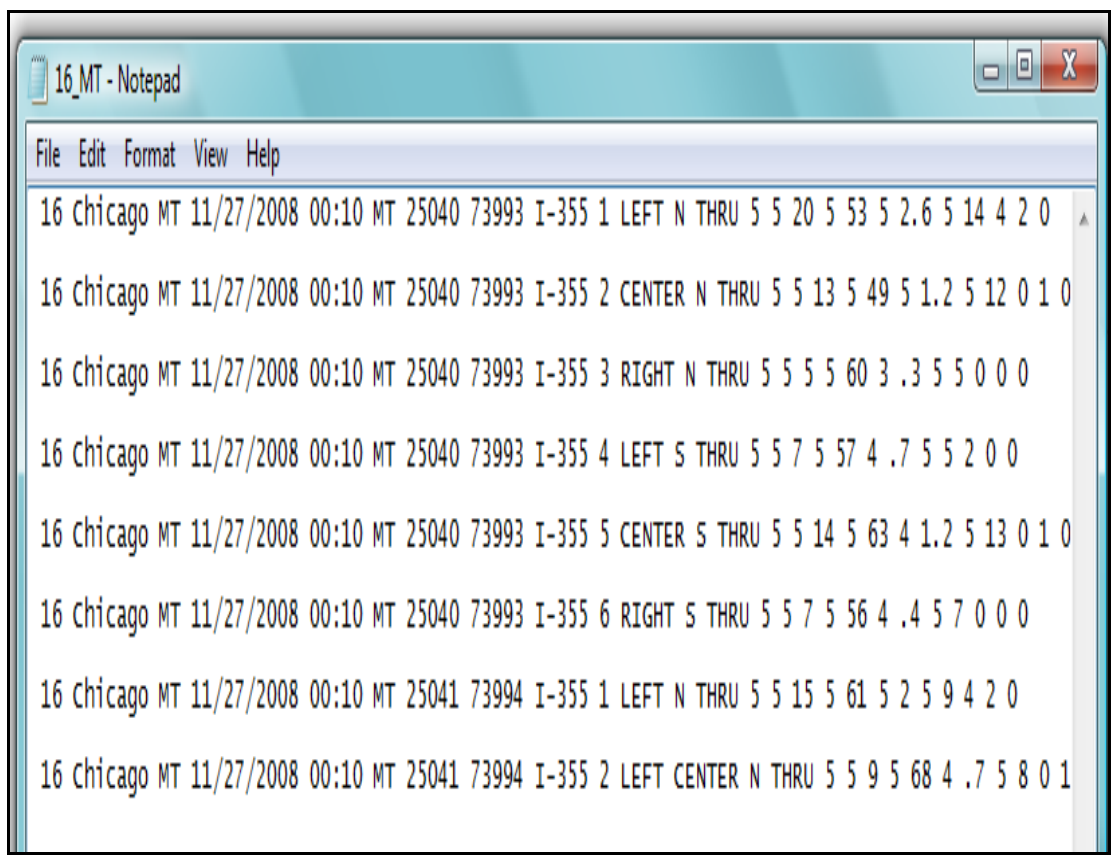

```
16_MT - Notepad
File Edit Format View Help
16 Chicago MT 11/27/2008 00:10 MT 25040 73993 I-355 1 LEFT N THRU 5 5 20 5 53 5 2.6 5 14 4 2 0
16 Chicago MT 11/27/2008 00:10 MT 25040 73993 I-355 2 CENTER N THRU 5 5 13 5 49 5 1.2 5 12 0 1 0
16 Chicago MT 11/27/2008 00:10 MT 25040 73993 I-355 3 RIGHT N THRU 5 5 5 5 60 3 .3 5 5 0 0 0
16 Chicago MT 11/27/2008 00:10 MT 25040 73993 I-355 4 LEFT S THRU 5 5 7 5 57 4 .7 5 5 2 0 0
16 Chicago MT 11/27/2008 00:10 MT 25040 73993 I-355 5 CENTER S THRU 5 5 14 5 63 4 1.2 5 13 0 1 0
16 Chicago MT 11/27/2008 00:10 MT 25040 73993 I-355 6 RIGHT S THRU 5 5 7 5 56 4 .4 5 7 0 0 0
16 Chicago MT 11/27/2008 00:10 MT 25041 73994 I-355 1 LEFT N THRU 5 5 15 5 61 5 2 5 9 4 2 0
16 Chicago MT 11/27/2008 00:10 MT 25041 73994 I-355 2 LEFT CENTER N THRU 5 5 9 5 68 4 .7 5 8 0 1
```

Figure 6. Extracted text stored in generated file 16_MT

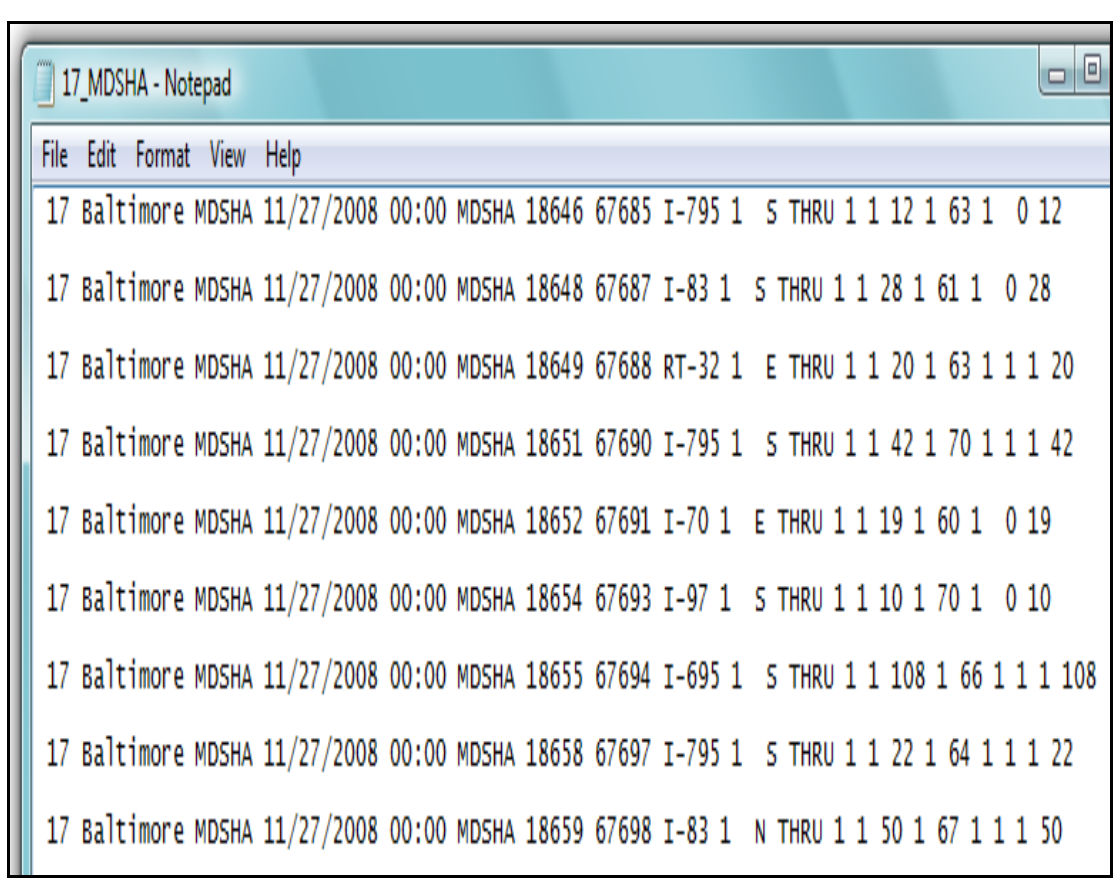

```
17_MDSHA - Notepad
File Edit Format View Help
17 Baltimore MDSHA 11/27/2008 00:00 MDSHA 18646 67685 I-795 1  S THRU 1 1 12 1 63 1  0 12
17 Baltimore MDSHA 11/27/2008 00:00 MDSHA 18648 67687 I-83 1  S THRU 1 1 28 1 61 1  0 28
17 Baltimore MDSHA 11/27/2008 00:00 MDSHA 18649 67688 RT-32 1  E THRU 1 1 20 1 63 1 1 1 20
17 Baltimore MDSHA 11/27/2008 00:00 MDSHA 18651 67690 I-795 1  S THRU 1 1 42 1 70 1 1 1 42
17 Baltimore MDSHA 11/27/2008 00:00 MDSHA 18652 67691 I-70 1  E THRU 1 1 19 1 60 1  0 19
17 Baltimore MDSHA 11/27/2008 00:00 MDSHA 18654 67693 I-97 1  S THRU 1 1 10 1 70 1  0 10
17 Baltimore MDSHA 11/27/2008 00:00 MDSHA 18655 67694 I-695 1  S THRU 1 1 108 1 66 1 1 1 108
17 Baltimore MDSHA 11/27/2008 00:00 MDSHA 18658 67697 I-795 1  S THRU 1 1 22 1 64 1 1 1 22
17 Baltimore MDSHA 11/27/2008 00:00 MDSHA 18659 67698 I-83 1  N THRU 1 1 50 1 67 1 1 1 50
```

Figure 7. Extracted text stored in generated file 17_MDSHA

First figure shows the progress of SRr running and rest of the snapshots show the contents segregated from the input text file. Figure 1 is a screen shot depicting the progress of SRr.

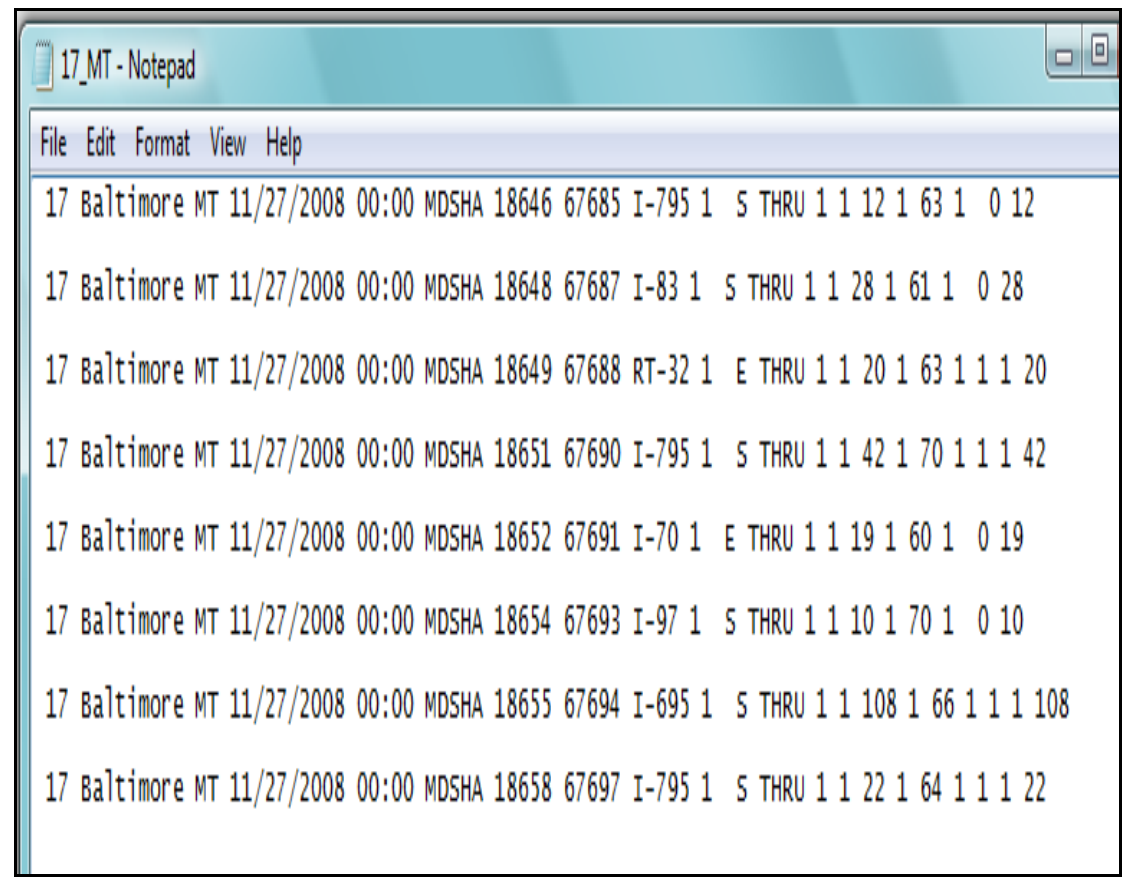

```
17_MT - Notepad
File Edit Format View Help
17 Baltimore MT 11/27/2008 00:00 MDSHA 18646 67685 I-795 1  S THRU 1 1 12 1 63 1  0 12
17 Baltimore MT 11/27/2008 00:00 MDSHA 18648 67687 I-83 1  S THRU 1 1 28 1 61 1  0 28
17 Baltimore MT 11/27/2008 00:00 MDSHA 18649 67688 RT-32 1  E THRU 1 1 20 1 63 1 1 1 20
17 Baltimore MT 11/27/2008 00:00 MDSHA 18651 67690 I-795 1  S THRU 1 1 42 1 70 1 1 1 42
17 Baltimore MT 11/27/2008 00:00 MDSHA 18652 67691 I-70 1  E THRU 1 1 19 1 60 1  0 19
17 Baltimore MT 11/27/2008 00:00 MDSHA 18654 67693 I-97 1  S THRU 1 1 10 1 70 1  0 10
17 Baltimore MT 11/27/2008 00:00 MDSHA 18655 67694 I-695 1  S THRU 1 1 108 1 66 1 1 1 108
17 Baltimore MT 11/27/2008 00:00 MDSHA 18658 67697 I-795 1  S THRU 1 1 22 1 64 1 1 1 22
```

Figure 8. Extracted text stored by in generated file 17_MT

Figure 2 shows the contents of the file generated by SRr. The name of the file is 1_IDOT. SRr extracts that data from the

## 4. CONCLUSION

SRr is a text mining tool performs mining operations on input text file and segregates the data lines based on composite attributes. We analyze the results generated by SRr and found that each file has distinct data stored in it based on its composite attributes. In this research paper the mining operation has been performed on an input text file consisting of traffic data but functioning of SRr is not confined till here. SRr is a customizable tool and can be applied to other applications as well where one reads an input file. In this research paper we have applied SRr on traffic data file and our next goal is to use SRr on file consisting genomics data and sequence analyzing subsequently. It is a rare and noble work we have done that Perl developed mining tool, SRr extracts the traffic data and segregates that for easy analyzing. We believe that our work will help communities, academia and industry and prove itself as a leader in text/data mining.